\documentclass[web, 10 pt, journal]{ieeetran}

\IEEEoverridecommandlockouts                              
                                                          
\usepackage{amsmath}
\usepackage{amssymb}
\usepackage{tabularx}
\usepackage{xfrac}
\usepackage[final]{pdfpages}
\usepackage{algorithm}
\usepackage[noend]{algpseudocode}
\usepackage{booktabs}
\usepackage[short]{optidef}
\usepackage{soul,xcolor}
\usepackage{siunitx}
\usepackage{import}
\usepackage{bm}
\usepackage{tablefootnote}
\usepackage{physics}
\usepackage{arydshln}
\usepackage{lipsum}
\usepackage{enumitem}

\usepackage{url}
\makeatletter
\let\NAT@parse\undefined
\makeatother
\usepackage[hidelinks]{hyperref}
\usepackage[capitalise,nameinlink,sort]{cleveref}
\usepackage{balance}

\usepackage{graphicx}
\usepackage{comment} 

\usepackage{cite}
\usepackage{mathtools,cuted}

\usepackage{caption}
\usepackage{subcaption}

\usepackage{nomencl}
\makenomenclature

\Crefname{equation}{Eq.}{Eqs.}
\Crefname{figure}{Fig.}{Figs.}
\Crefname{tabular}{Tab.}{Tabs.}
\Crefname{line}{L.}{L.}

\renewcommand{\nompreamble}{The next list describes several symbols that we used throughout the paper and their explanation.}

\newlength\figureheight
\newlength\figurewidth

\newcommand{\Pkp}{P_{\text{kp}}} 
\newcommand{\Vkp}{V_{\text{kp}}} 
\newcommand{\Vki}{V_{\text{ki}}} 
\newcommand{\Cp}{C_{\text{p}}} 
\newcommand{\Cv}{C_{\text{v}}} 
\newcommand{\Vff}{V_{\text{ff}}} 
\newcommand{\Aff}{A_{\text{ff}}} 

\newcommand{\suggestion}{x^*}

\newcommand{\threshold}{c}
\newcommand{\expansionop}{g_\epsilon^t}
\newcommand{\task}{x_{\tau}}
\newcommand{\optimizable}{x_\text{opt}}

\newcommand{\mypar}[1]{\textbf{#1.}}

\newcommand{\goose}{\textsc{GoOSE}}

\newboolean{include-notes}
\setboolean{include-notes}{true}

\newcommand{\matteo}[1]{\ifthenelse{\boolean{include-notes}}{{\color{orange} \textbf{Matteo}: #1}}{}}
\newcommand{\christopher}[1]{\ifthenelse{\boolean{include-notes}}{{\color{green} \textbf{Christopher}: #1}}{}}
\newcommand{\alisa}[1]{\ifthenelse{\boolean{include-notes}}{{\color{red} \textbf{Alisa}: #1}}{}}
\newcommand{\andreas}[1]{\ifthenelse{\boolean{include-notes}}{{\color{blue} \textbf{Andreas}: #1}}{}}
\newcommand{\raamadaas}[1]{\ifthenelse{\boolean{include-notes}}{{\color{blue} \textbf{Raamadaas}: #1}}{}}

\newcommand{\R}{\mathbb{R}}

\newcommand{\X}{\mathcal{X}}
\newcommand{\SafeSet}{\X_{t}}
\newcommand{\OptimisticSafeSet}{\X^{opt}_{t}}
\newcommand{\SafeSeed}{\X_{0}}
\newcommand{\ExpanderSet}{W_{t}}

\newcommand{\lcb}{\mathrm{lcb}}
\newcommand{\ucb}{\mathrm{ucb}}

\newcommand{\argmin}{\operatornamewithlimits{argmin}}

\newcommand{\sigmav}{\sigma_v}
\newcommand{\likelihood}{\sigma_{n}}
\newcommand{\lengthscale}{L}
\newcommand{\prior}{\mu_0}
\newcommand{\kernel}{\kappa}

\begin{document}


\title{Adaptive Bayesian Optimization for \\ High-Precision Motion Systems 
}
\author{Christopher K{ö}nig$^{1}$, Raamadaas Krishnadas$^{1}$, Efe C. Balta$^{1}$, and Alisa Rupenyan$^{2}$
	\thanks{This project was funded by the Swiss Innovation Agency, grant Nr. 46716, and by the Swiss National Science Foundation under NCCR Automation, grant Nr. 180545. \newline
		$^{1}$ Control and Automation Group, Inspire AG, Z\"{u}rich, Switzerland. \newline
		$^{2}$ ZHAW Centre for Artificial Intelligence, Z\"{u}rich University for Applied Sciences, Z\"{u}rich, Switzerland. \newline Corresponding author: E. C. Balta: \texttt{efe.balta@inspire.ch} }}

\maketitle


\begin{abstract}
 Controller tuning and parameter optimization are crucial in system design to improve closed-loop system performance. Bayesian optimization has been established as an efficient model-free controller tuning and adaptation method. However, Bayesian optimization methods are computationally expensive and therefore difficult to use in real-time critical scenarios. In this work, we propose a real-time purely data-driven, model-free approach for adaptive control, by online tuning low-level controller parameters. We base our algorithm on GoOSE, an algorithm for safe and sample-efficient Bayesian optimization, for handling performance and stability criteria. We introduce multiple computational and algorithmic modifications for computational efficiency and parallelization of optimization steps. We further evaluate the algorithm's performance on a real precision-motion system utilized in semiconductor industry applications by modifying the payload and reference stepsize and comparing it to an interpolated constrained optimization-based baseline approach.
\end{abstract}
\renewcommand{\abstractname}{Note to Practitioners}	
\begin{abstract}
This work is motivated by developing a comprehensive control and optimization framework for high-precision motion systems. Precision motion is an integral application of advanced mechatronics and a cornerstone technology for high-value industrial processes such as semiconductor manufacturing. The proposed method framework relies on data-driven optimization methods that can be designed by prescribing desired system performance. By using a method based on Bayesian Optimization and safe exploration, our method optimizes desired parameters based on the prescribed system performance. A key benefit is the incorporation of input and output constraints, which are satisfied throughout the optimization procedure. Therefore, the method is suitable for use in practical systems where safety or operational constraints are of concern. Our method includes a variable to incorporate contextual information, which we name the task parameter. Using this variable, users can input external changes, such as changing step sizes for a motion system, or changing weight on top of the motion system. We provide two parallel implementation variants of our framework to make it suitable for run-time operation under context changes and applicable for continuous operation in industrial systems. We demonstrate the optimization framework on simulated examples and experiment on an industrial motion system to showcase its applicability in practice. 
\end{abstract}
\section{Introduction}
High-performance mechanical positioning systems used in high-yield industries such as semiconductor manufacturing rely on precision motion systems. Due to the level of precision required in such applications, closed-loop control is necessary, especially in the context of run-time disturbance, mechanical nonlinearities, and context shifts in practice. 
Consequently, precision motion control has been a well-studied topic in mechatronics systems~\cite{tan2007precision}.

Adaptive control approaches present a compelling alternative to robust controllers, particularly in high-performance precision motion applications, where tracking performance is crucial.
The investigation into addressing uncertainties inherent in dynamics and facilitating adaptation has been studied in classical adaptive control literature via methods such as Model Reference Adaptive Control (MRAC)~\cite{Chowdhary2015,Grande2014}. Gaussian processes (GP) have also found utility in the context of modeling the outputs of nonlinear systems within a dual controller paradigm \cite{Sbarbaro2005}, where the coupling of system states and inputs is encapsulated within the covariance function of the GP model. Learning system dynamics using a $\mathcal{L}_1$-adaptive control framework has been additionally demonstrated~\cite{gahlawat20a,Fan2019}.

Rather than pursuing the modeling or learning of system dynamics, an alternate approach involves characterizing the system through its observed performance, directly extracted from measurement data. Subsequently, optimization of the low-level controller parameters can be undertaken to meet the desired performance criteria, as demonstrated in the context of motion systems \cite{Berkenkamp2, Duivenvoorden, khosravi2020cascade,Khosravi2020, khosravi2022safety}. However, the application of such model-free methodologies to continuous adaptive control remains limited, primarily due to challenges associated with maintaining stability and safety under the influence of disturbances and system uncertainties, along with the computational complexity of such methods. 

Improving the performance of motion systems requires maintaining good positioning accuracy during motion while ensuring a short settling phase, which is possible by eliminating run-time vibrations and other disturbances. Common methods for motion control include gain scheduling, LPV control approaches, and tuning of low-level controller parameters. However, gain scheduling and LPV models rely on a static view of the system's performance, and are usually tuned to fixed step sizes.
Similarly, parameter tuning, which is often implemented via automated routines, suffers from the same problem of lacking adaptivity and flexibility to changes in the system.
In the recent literature, run-to-run adaptation of the control parameters in mechanical systems has been demonstrated using an efficient Bayesian optimization-based approach, which prevents instabilities while maximizing tracking performance~\cite{koenig2021}. However, the approach is limited to slow updates and simple systems, due to its associated computational load.

\subsection{Contributions}
This paper proposes an efficient data-driven algorithm for run-to-run control parameter adaptation, suitable for real-time applications, and showcases its efficacy for high-precision motion systems.
We summarize our main contributions below.
\begin{enumerate}[leftmargin=*]
    \item We provide a safe and adaptive control algorithm for real-time applications based on parallelization of the optimization via multiprocessing. We accomplish this by separating the optimization into an active- and passive phase, between which the algorithm switches, based on event-triggered criteria.
    \item Building on the previously introduced \goose{} algorithm for data-driven optimization, we present a computationally efficient implementation, by stripping the algorithm of its discretization of the domain.
    \item We show the effectiveness of our approach on a real nanometer precision motion system, a crucial instrument in the semiconductor industry.
\end{enumerate}

\subsection{Related work and paper organization}
\label{sec:lit}

\mypar{Adaptive control parameter tuning for motion systems} 
An adaptive control method with performance guarantees in the presence of model uncertainty is Model Reference Adaptive Control (MRAC) \cite{zhang2017review}, which can be enhanced by modeling the uncertainties in the reference model, as demonstrated in  \cite{Grande2014} for a quadrotor system. 
Additionally, adaptive robust control methods for precision motion have been developed for dealing with parametric uncertainties~\cite{xu2001adaptive}. 
However, such adaptive controllers require run-time computation and possible changes to the controller architecture, which may not be practical in certain high-speed applications.

Linear parameter varying gain scheduling methods to adapt to sudden changes in the system dynamics of motion systems have previously been designed and studied by \cite{Liu2015} for unmanned surface vehicles. The repetitive nature of motion in mechatronics systems has been exploited in various iterative learning control-based approaches (ILC), for example, combined with mismatch modeling to reduce the tracking error \cite{balta2021learning}, or with feed-forward compensation of acceleration, jerk, and snap terms as demonstrated in \cite{Dai2021}. While ILC approaches provide an excellent error reduction for repetitive motion, they require additional development to deal with changing trajectories and disturbances.  Depending on the system, this can be overcome by describing parametrically the reference and the plant's dynamics  \cite{Hayashi2019} and connecting this parametric description to the ILC update law. 

\mypar{BO for controller tuning}
Bayesian Optimization (BO) has found applications in refining a variety of controller types designed for high-speed precision motion systems. BO-driven tuning aimed at performance enhancement has been effectively showcased in the context of cascade controllers governing linear axis drives \cite{khosravi2020cascade,rupenyan2021performance}. This approach involves leveraging data-derived performance metrics to deliberately enhance traversal time and tracking accuracy while reducing vibrations. In addition, models describing the noise in the system can be added to improve performance and safety \cite{koenig2023risk}. 

In the realm of model predictive control (MPC), an alternative strategy emerges by selecting the prediction model to optimize closed-loop performance. Selecting the prediction model is in contrast to robust control design methods of designing controllers for worst-case scenarios. The controller optimization involves tuning parameters within the MPC optimization objective to achieve optimal performance \cite{rupenyan2021performance}. While BO is efficient for finding good controller parameters before deploying a system in operation, it has been less explored for adaptive tuning during operation. The reasons for the lack of such studies are twofold. Firstly, it requires simplifying assumptions and prior knowledge on the nature of the expected changes in a system \cite{brunzema2022controller}, and, secondly, it requires a mechanism for continuously reducing data from past time instances \cite{koenig2021} which might reduce the predictive capabilities of the approach.

\mypar{Safe BO}
To maintain optimal closed-loop performance for systems subject to variations, controllers must dynamically adapt their parameters online, all while avoiding configurations that might lead to instability. The need for online adaptation is particularly pronounced for high-precision motion systems, where even a single iteration marked by excessive vibrations is deemed unacceptable during continuous operation. Consequently learning the effect of the cascaded controller parameters under safety and stability constraints is necessary to ensure that only safe parameters are explored~\cite{koenig2020}.
One approach to address this challenge involves the utilization of the SafeOpt algorithm introduced in~\cite{Sui}. Notably, while this algorithm is effective in ensuring safety, its exploration strategy may be inefficient in practice~\cite{berkenkamp2021bayesian}. To improve the computational and sample efficiency of SafeOpt, optimization-based search approaches have been recently demonstrated~\cite{zagorowska2023efficient}.

An alternative was explored in~\cite{fiducioso2019safe}, where the safe set is not actively expanded. While the method in~\cite{fiducioso2019safe} might entail a trade-off with optimality, it aligns well with the considered application.
Additionally, a distinct alternative for controller tuning involves the incorporation of safety-related log barriers within the cost function, as exemplified by the work presented in~\cite{khosravi2022safety}. This innovation aims to curtail constraint violations, as compared to approaches grounded in Bayesian optimization with inequality constraints, such as the one outlined in~\cite{Gardner}.

\mypar{GoOSE}
An efficient strategy for facilitating safe exploration is introduced in the work of GoOSE \cite{turchetta2019safe}, where the overarching objective is to ensure that all inputs introduced to the system adhere to an unknown yet observable constraint. In the context of controller tuning, GoOSE unifies time-varying Gaussian process bandit optimization \cite{bogunovic2016time}, multi-task Gaussian processes \cite{swersky2013multi}, and an efficient safe set search mechanism grounded in particle swarm optimization~\cite{wang2018particle}. The result is a cohesive framework that proves instrumental in the adaptive tuning of motion systems \cite{koenig2021}.
However, a noteworthy limitation emerges due to the discretization of the input space of the expanders. This limitation renders the algorithm impractical for scenarios characterized by high-dimensional input and task spaces, or instances involving substantial amounts of data within real-time online applications.

The rest of the paper is organized as follows. 
\Cref{sec:method} presents the proposed method with implementation details. \Cref{sec:case_study} presents numerical and experimental studies to demonstrate the effectiveness of the proposed method. \Cref{sec:conclusion} provides concluding remarks and future directions.

\section{ Problem setting and Preliminaries }\label{sec:method} 

\label{sec:mogoose}

Consider a closed-loop control system of a possibly nonlinear plant. Let $\optimizable \in \X_\text{opt}$ be a vector of controller parameters, such as PID controller gains or feedforward gains in the closed loop. 
Furthermore, let $\task \in \X_{\task}$ be the context parameters of the system influencing the behavior of the system (e.g. temperature, payload, reference parameters). 
The complete input can be written as $x = [\optimizable, \task]^{T} \in \X$. The parameters $\optimizable$ should be continuously optimized online and in real-time, i.e., while the system is operating, such that the overall closed-loop performance of the system is optimized for the current context ${\task}_{,t}$, while stability is ensured at any time.
This can be expressed by the constrained optimization problem:
\begin{equation}
    \begin{aligned}
    \min_{x \in \X} \{ f(x)~|~ q(x) \leq \threshold \And \task = {\task}_{,t} \},
\end{aligned}
\label{eq:constrained_bo}
\end{equation}
under the feasibility assumption $\{x \; | \; q(x) \leq \threshold \} \neq \emptyset$, for a given $\textstyle{c \in \R}$ at any possible context $\task \in \X_{\task}$.
The unknown cost function $f(x)$ and the constraint or safety metric $q(x)$ here are expensive-to-evaluate functions, extracted from experimental data, that should be learned on-the-fly.

\begin{figure}
\centering
    \includegraphics[width=0.85\columnwidth]{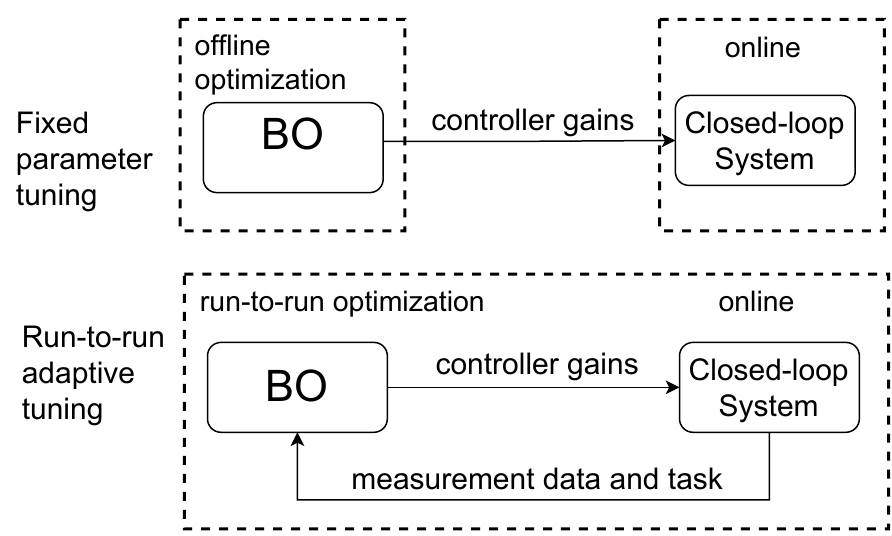}
    \caption{The BO-based run-to-run controller tuning setting considered in this work.}    
    \label{fig:BO_scheme} 
\end{figure}

The setting of the adaptive controller tuning framework presented in this paper is given in \cref{fig:BO_scheme}.
Traditional BO approaches in the literature aim to optimize the controller parameters in an offline optimization step, which is then deployed for online usage. 
In this work, we continuously update the controller parameters in a run-to-run fashion using the measurement in each run with the goal of solving~\eqref{eq:constrained_bo}.
To ensure desirable transient operation and safety constraints, we base our framework on the \goose{} method, presented below.
The main elements of the algorithm enabling continuous safe optimization are built on \goose{} developed in~\cite{koenig2021} and multi-task optimization. 

\subsection{Probabilistic models}
To reason under uncertainty about the unknown objective $f(x)$ and safety metrics $q(x)$, we model each as a gaussian process - $GP^{f}$ and $GP^{q}$, characterised by kernel functions $\kernel^{f}$ and $\kernel^{q}$ as well as priors $\prior^{f}$ and $\prior^{q}$, respectively. Note that in the following we have omitted the $f$ and $q$ indices for ease of notation as the expressions are identical for both. Posterior GP mean and variance denoted by $\mu_t(\cdot)$ and $\sigma^2_t(\cdot)$, respectively, are computed based on the previous measurements $y_{1:t} = [y_1, \dots, y_t]^\top$ and a given kernel $\kappa(\cdot,\cdot)$, where $t$ is the number of observations :
\begin{align}
    \begin{aligned}
        \mu_{t}(x) &= \kernel_t(x)^T(K_t + \sigma_{n}^{2} I)^{-1} (y_{1:t} - \prior(x)), \\
        \sigma_{t}^{2}(x) &= \kernel(x, x) - \kernel_t(x)^T(K_t + \sigma_{n}^{2} I)^{-1} \kernel_t(x), 
    \end{aligned}
    \label{eq:GP_posterior}
\end{align}
where $I \in \R^{t \times t}$ is the identity matrix, $\sigma_{n}$ is the noise variance of the GP, $K_t$ is the covariance matrix with $(K_t)_{i,j} = \kernel(x_i, x_j)$, $\ \kernel_t(\cdot)^T = [\kernel(x_1, \cdot), .., \kernel(x_t, \cdot) ]^T$.
We follow \cite{koenig2021}, assuming that the functions $f(x)$, $q(x)$ belong to a reproducing kernel Hilbert space (RKHS)  associated with their respective kernels, and have bounded RKHS-norms. This assumption enables building well-calibrated confidence intervals over the two funcitons. To build the confidence intervals, we define the lower and upper confidence bounds $\lcb^{q}_{t}(x), \, \ucb^{q}_{t}(x)$ as follows:
\begin{align}
        \lcb_{t}(x) &= \mu_{t}(x) - \beta^{f/q} \sigma_{t}(x), \\
        \ucb_{t}(x) &= \mu_{t}(x) + \beta^{f/q} \sigma_{t}(x) .  \label{eq:lcb_ucb} 
\end{align}
The corresponding parameters $\beta^f$, $\beta^q$
 can be individually adjusted to ensure the validity of confidence bounds \cite{kirschner2018information} and balance exploration vs. exploitation (further referenced collectively as $\beta$ parameters). \looseness-1

\subsection{Online adaptation under safety constraints}
In this work the underlying assumption is that for each measurement the task $\task$ can be quantified.
The online adaption is enabled by using multi-task Gaussian Processes, where each measurement can be associated with a context and its corresponding task parameter.

The kernel of a multi-task GP is of the form $k_\text{multi}((\optimizable,\task),(\optimizable',\task')) = k_{\tau}(\task,\task')\otimes k(\optimizable,\optimizable')$, where $\otimes$ denotes the Kronecker product. This kernel decouples the correlations in objective values along the input dimensions, captured by $k$, from those across tasks, captured by $k_\tau$ \cite{Swersky2013}. Before the initialization of BO, \goose{} assumes a known non-empty safe seed $\SafeSeed$~\cite{koenig2021}. 
For online adaption using different tasks/contexts, $\SafeSeed$ needs to be safe and fulfill the constraint(s), i.e., $q(\SafeSeed) < c$ for all tasks. The safe seed $\SafeSeed$ is also used as a backup solution for the BO if the set of safe points within the current task $\task$ is empty, e.g. if the task changes drastically and no data is available to the GPs within a neighborhood of the new task parameters. In this case external prior knowledge of the operator is used.

In \cite{koenig2021}, \goose{} divides the optimization domain into two subsets: \emph{the pessimistic safe set} $\SafeSet$ and \emph{the optimistic safe set} $\OptimisticSafeSet$. $\SafeSet$ contains the set of points classified as safe while $\OptimisticSafeSet$ contains the points that potentially could be safe at iteration $t$. Formally,  
    \begin{align}
        \SafeSet &= \{x \in \X \; | \; \ucb^{q}_{t}(x) \leq \threshold \},  \\
        \OptimisticSafeSet &= \{ x \in \X \; | \; \exists \bar{x} \in \ExpanderSet, \text{ s.t. } \expansionop(\bar{x},x) > 0  \}\,,
        \label{eq:pessimistic_safe_set_condition}
    \end{align}
where the subscript $t$ indicates the iteration of the Bayesian optimization, and the superscript $q$ marks the constraint $q(x)$.
The \emph{set of expanders} $\ExpanderSet \subset \SafeSet$ corresponds to the periphery of the safe set, such that $\textstyle{\forall \; x \in W_{t}}$ it holds true that $\textstyle{|\ucb^{q}_{t}(x) - \lcb^{q}_{t}(x)| > \epsilon}$. Furthermore, $\expansionop(\bar{x},x)$ is the noisy expansion operator taking values of $0$ or $1$, defined as \looseness-1
\begin{equation}
\expansionop(\bar{x},x) =\mathbb{I}\left[\lcb_{t}^{q}(\bar{x})+\left\|\mu_{t}^{\nabla}(\bar{x})\right\|_{\infty} d(\bar{x}, x)+\epsilon \leq \threshold \right] \,,
\label{eq:noisy_expansion_operator}
\end{equation}
for some $\textstyle{\epsilon > 0}$ uncertainty threshold relating to the observation uncertainty of $q(x)$. $\left\|\mu_{t}^{\nabla}(\bar{x})\right\|$ is the mean of the posterior over the gradient of the constraint function, and $d(\cdot, \cdot)$ is the distance metric associated with Lipschitz continuity of $q(x)$.

In the original \goose{} formulation in \cite{turchetta2019safe}, the sets $\SafeSet$ and $\OptimisticSafeSet$ need to be explicitly calculated during each BO iteration, following discretization of the whole optimization domain. The proposed definitions in \eqref{eq:pessimistic_safe_set_condition} reduce the explicit calculation of the sets $\SafeSet$ and $\OptimisticSafeSet$ to criteria that can be checked for each candidate $x$. However, discretizing the domain to build the set of expanders $\ExpanderSet$ cannot be avoided, slowing down the algorithm for high dimensional domains or upon continuous optimization.

\section{\textsl{Modified \goose{}}: Safe Continuous Goal-Oriented BO}
To enable fast online optimization, we remove the expander search from the \goose{} algorithm of \cite{koenig2021}, without losing its safe expansive behaviour due to thoughtfully chosen hyperparameters, $\beta$ parameters and prior settings coming from prior knowledge of the objective function $f(x)$ and the constraint $q(x)$. 
In this work, we use the squared exponential kernel
\begin{equation}
    \kernel(x,x') = \sigmav^2 \exp{-\frac{(x-x')^T(x-x')}{2\lengthscale}},
    \label{eq:squared_exponential_kernel}
\end{equation}
where $\sigmav$ is the prior standard deviation of the GP and $\lengthscale = \text{diag}(l_1, ..., l_m)$ is the diagonal matrix of the lengthscales for the corresponding dimensions of the input $x$.
The modified \goose{} algorithm works with various other kernels e.g. Matérn kernels.
As in \cite{koenig2021}, the acquisition function for selecting the next point is the lower confidence bound search
\begin{equation}
    x_t = \argmin_{x \in \SafeSet} \lcb^f_t(x).
    \label{eq:lcb_search}
\end{equation}

We introduce a method to make the optimizers of the modified GoOSE algorithm expansive and safe and therefore loose the need of explicit expanders.
For this purpose we extend the algorithm of \cite{koenig2021} by the additional assumption on $f(x)$, lower-bounding the cost by a real constant value:
\begin{equation}
        \{\exists \xi \in \R | \; f(x) \geq \xi, \forall x \in \X\}.
    \label{eq:assumption_mogoose}
\end{equation}

This assumption can be easily satisfied for many practical applications, e.g. using quadradic or other non negative costs. 
In Section~\ref{sec:case_study}, we demonstrate a non negative $f(x)$, which fits well with our optimization objective.
For safe expansive behaviour of the optimizers, the priors $\prior^{f}(x)$ and $\prior^{q}(x)$, the hyperparameters of the prior standard deviation $\sigmav^f$ and $\sigmav^q$, and the $\beta$ parameters $\beta^f$ and $\beta^q$ have to be tuned such that
\begin{equation}
    \begin{aligned}
        \prior^{q}(x) + \beta^q\sigmav^q &> \threshold \quad \forall x \in \X, \\
        \prior^{f}(x) - \beta^f\sigmav^f &\leq \xi \quad \forall x \in \X, \\
    \end{aligned}
    \label{eq:hyperparameter_setting_mogoose}
\end{equation}
meaning that prior to adding any data to the GPs at $t=0$, all $x \in \X$ are outside of the safe set $\SafeSet$, while the lower bound for unknown points $x \in X$ is less or equal to the absolute possible minimum of the cost function $f(x)$ (following from \eqref{eq:assumption_mogoose} and \eqref{eq:hyperparameter_setting_mogoose}).
Intuitively the first condition ensures safety in the presence of uncertainty and the second condition ensures expansion in the presence of uncertainty, by providing a candidate $x \in \X$ for the acquisition function (see \eqref{eq:lcb_search}), that is always better than the current best observation $\min (y_{1:t}) \geq \xi$.

\subsection{Implementation of the Modified \goose{}}

\begin{algorithm}
\caption{Modified \goose{}} 
\label{alg:adaptive_control_goose}
\textbf{Input}: Safe seed $\SafeSeed$, $f\sim \mathcal{GP}^f(\prior^f,\kernel^f;\likelihood^f, \beta^f)$, $q\sim \mathcal{GP}^q(\prior^q,\kernel^q;\likelihood^q, \beta^q)$, $\task = {\task}_{0}$\, 

$\prior^f$, $\prior^q$, $\likelihood^f$, $\likelihood^q$, $\beta^f$, $\beta^q$ tuned according and \ref{eq:assumption_mogoose} and \ref{eq:hyperparameter_setting_mogoose}
   \begin{algorithmic}[1]
       \While {machine is running}
            \While {termination criteria is false} \Comment{Active phase}
                \State{$\bm{x_w} \gets ({\optimizable}_{,1:t-1}, {\task}_{,t})$}
                \State{${\SafeSet}_{w} \gets \{x \in \bm{x_w}:\ucb^q_t(x) \leq\ c\}$} \label{alg:SafeSetCalc}
                \State{$\suggestion_t \gets\! \texttt{PSO}(\mathrm{init}\!:\!{\SafeSet}_{w},\;\mathrm{acq}\!:\!\lcb^f_t,\;\mathrm{s.t.}\!:\!\ucb^q_t \leq c)$} \label{alg:pso_activephase}
                \State{Evaluate $f(\suggestion),q(\suggestion)$}
                \If{length($\bm{x_w}$) $\geq$ $\text{data}_{\text{limit}}$}
                {
                \newline \hspace*{5em} remove oldest data point from \newline
                \hspace*{5em} $\mathcal{GP}^f$ and $\mathcal{GP}^q$
                }
                \EndIf
                \State{Update $\mathcal{GP}^f_t$ and $\mathcal{GP}^q_t$ with $\suggestion,\;f(\suggestion),\;g(\suggestion)$} \label{alg:GP_updates}
                \State{Update ${\task}_{,t}$}
            \EndWhile
            \While {restart criteria is false} \Comment{Passive phase}
                \State{$\bm{x_w} \gets ({\optimizable}_{,1:t-1}, {\task}_{,t})$}
                \State{${\SafeSet}_{w} \gets \{x \in \bm{x_w}:\ucb^q_t(x) \leq\ c\}$}
                \State{$\suggestion_t \gets\! \texttt{PSO}(\mathrm{init}\!:\!{\SafeSet}_{w},\;\mathrm{acq}\!:\!\ucb^f_t,\;\mathrm{s.t.}\!:\!\ucb^q_t\! \leq\! c)$} \label{alg:pso_passivephase}
                \State{Evaluate $f(\suggestion),q(\suggestion)$}
                \State{update ${\task}_{,t}$}
            \EndWhile
        \EndWhile
    \end{algorithmic}
\end{algorithm}
The modified \goose{} is presented in \cref{alg:adaptive_control_goose}. It consists of an active and a passive phase. In the active phase, the GPs are updated with data obtained by the measurement. The model is updated until it has sufficiently learned the system performance and is terminated with a termination criterion. The termination criterion should represent the convergence of the optimization. In the experiments of \cref{sec:case_study} we use a set of 30 additional data points added to the GPs. At each iteration $t$, it is checked which of the evaluated ${\optimizable}_{,1:t-1}$ are safe for the applied task ${\task}_{,t}$, line \ref{alg:SafeSetCalc}. This is the safe set which is given as an initialization seed for the particle swarm optimization ($\texttt{PSO}$) of the acquisition function, line \ref{alg:pso_activephase}. In case this set is void, the predefined safe seed is used as the safe set. In the active phase, the acquisition function is set to equal to $\lcb$. The $\texttt{PSO}$ provides an optimizer $x^*$ which is applied to the system. The obtained measurement is used for updating the GPs, line \ref{alg:GP_updates}. In the passive phase, the learned models $\mathcal{GP}^f$ and $\mathcal{GP}^q$ are used to find the optimum input for the specific task by minimizing the $\ucb$, line \ref{alg:pso_passivephase}. The passive phase operates until the restart criterion is triggered, which is the case when the models are not accurate enough and need to be updated. In our experiments in \cref{sec:case_study}, the restart criterion is in case of a constraint violation or new task detection. 

\subsection {Parallel computation schemes for fast optimization}
Depending on the use case, one or more task/context parameters may change rapidly. Serially calculating the optimizers might not be sufficient to keep pace with the changing context.
To be able to anticipate the next optimization step in (near) real-time, we need to develop strategies to move the whole frontier of points given by all possible contexts (or a large set of them) and to have predictions for the upcoming iterations for the whole frontier.

\begin{figure}
\centering
  \begin{subfigure}[h]{1.0\columnwidth}
    \centering    \includegraphics[width=\textwidth]{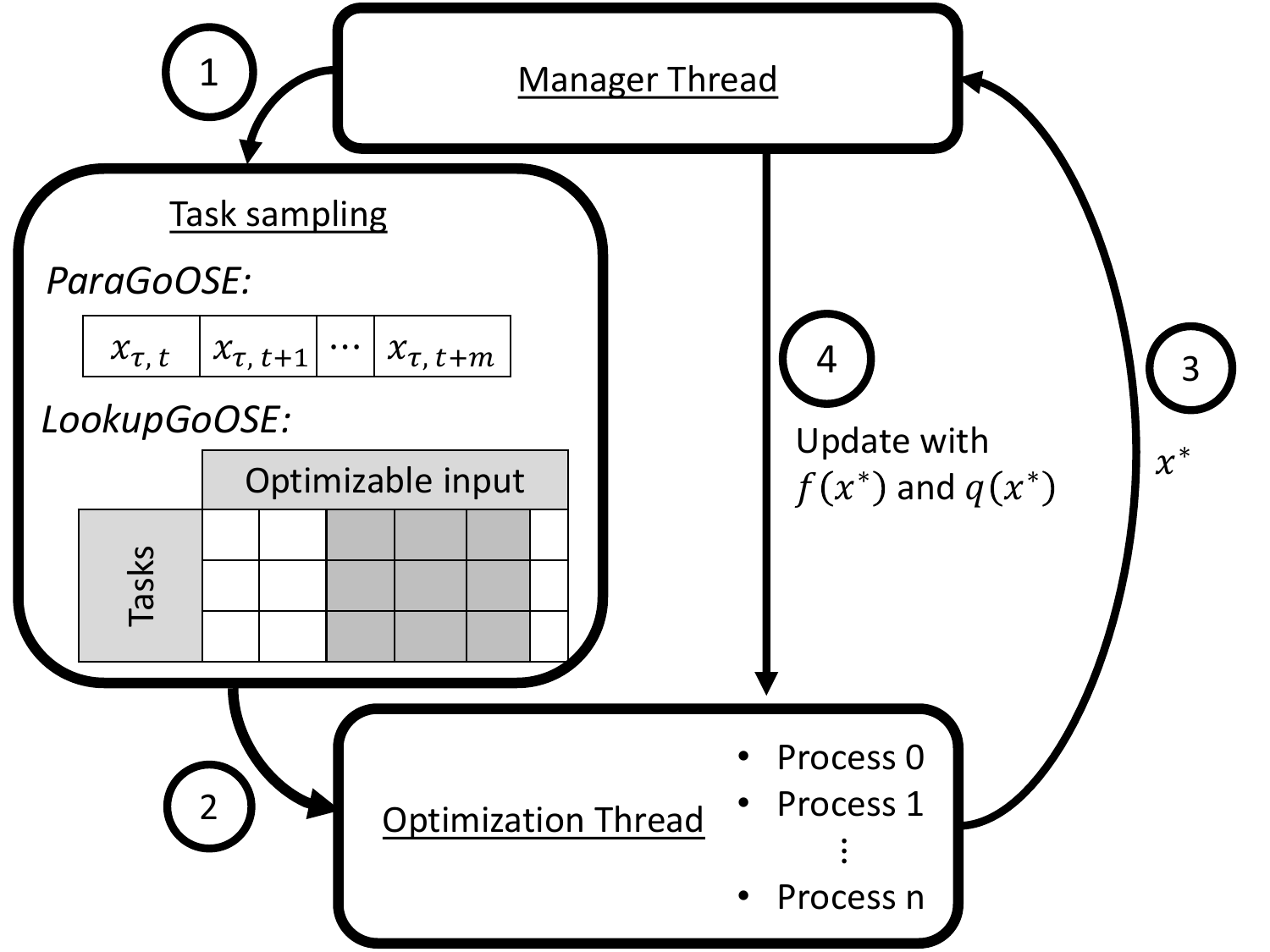}
  \end{subfigure}
  \caption{Parallel computation scheme to run the modified \goose{}.
    1. ParaGoOSE: A horizon of the next m tasks is sampled  and fed to the queue.
    LookupGoOSE: Corresponding to the next task, a neighborhood of tasks within the grid is selected and fed to the queue.
    2. The optimization processes pick the entries in the task queue and calculate the corresponding optimizers/optima according to algorithm \ref{alg:adaptive_control_goose}
    3. The optimizers/optima are returned to the manager thread to be executed.
    4. The $\mathcal{GP}^f$ and $\mathcal{GP}^q$ are updated with $f(x^*)$ and $q(x^*)$.}
  \label{fig:parallel_algorithm_scheme} 
\end{figure}

In \cref{fig:parallel_algorithm_scheme}, we present the overall parallel computation scheme proposed in this work.
In the first method, (Para\goose{}) (see \cref{fig:parallel_algorithm_scheme}), the optimization algorithm of \cref{alg:adaptive_control_goose}) is calculated $m$ times on $m$ different processes. A receding horizon of the next $m$ observations is selected/predicted by the manager thread and fed to the queue of tasks. The prediction requires access in advance to the next tasks. The $m$ processes are selected, removed, and optimized for the tasks in the queue until it is empty in a first come, first serve manner. Note that the number of processes can be less than the horizon length. Synchronization of the processes makes sure that the optimizations on all processes are in the same phase and have the same amount of data points added to the GPs. If the optimization is too slow and a measurement happens before the process is updated, the measurement gets ignored. Should there be no corresponding optimizer/optimum for a given task, due to the optimization being too slow, the predefined safe input $\SafeSeed$ is used.

The second method (Lookup\goose{}) uses a discretized set of task parameter combinations $\X_{\task}$. For each task in $\X_{\task}$, the optimum/optimizer is calculated and saved in a lookup table of optima/optimizers, which is initialized with the predefined safe input $\SafeSeed$.
Multiple optimization processes update the optima/optimizer corresponding neighborhood of the applied task ${\task}_{,t}$ of the new data within the discretized set $\X_{\task}$ added to a queue. The neighborhood is defined as: 
\begin{equation}
    {\task \in \X_{\task} : |\task - {\task}_{,t}| < k\Delta\task, \; k>0},
\end{equation}
where $\Delta\task$ is the discretization parameter of $\X_{\task}$.
When the phase switches the complete grid of optima/optimizers has to be updated.
Equivalent to Para\goose{}, the optimization processes are synchronized to be in the same state and the queue of tasks is emptied in a first come, first served manner.

One advantage of Para\goose{} is that the task for which the optimizer/optimum was calculated, corresponds exactly to the task predicted $n$ iterations ago, compared to Lookup\goose{}, resulting in more precise optimizers/optima given that the prediction of the task is correct. Due to only a single task update for all processes every iteration the method is also computationally less expensive than Lookup\goose{} and/or converges faster due to a lower ignore rate of the measurements.
The disadvantage of Para\goose{} compared to Lookup\goose{} is that a $n$ step horizon of the task needs to be computed. Unexpected changes in the task could lead to optima/optimizers calculated for the wrong context, which could lead to suboptimal and unsafe optimizers. Furthermore, each optimization result is calculated on the information level of $n$ iterations in the past. 

Lookup\goose{} requires no knowledge of the next $\task$. Furthermore, its asynchronous update of the optima/optimizer makes it less dependent on computational resources.
On the downside, new data can only be added to the GPs once the update of $\X_{\task}$ is finished. Additionally, data acquired within an update step is ignored. 
Although the same strategy is applied to Para\goose{}, this happens only occasionally due to the much lower computational cost of having only a single new task to optimize for each measurement.
Therefore the adaptation rate of Lookup\goose{} is slower than that of Para\goose{}.
When switching from Active Phase to Passive Phase or vice versa (see \cref{alg:adaptive_control_goose}) Lookup\goose{} needs to optimize for the whole task grid $\X_{\task}$, due to the change of the acquisition function. This update is computationally expensive and another disadvantage of Lookup\goose{} compared to Para\goose{}.

\section{Case Study}
\label{sec:case_study}
In this section, we demonstrate the proposed approach, first on a numerical simulation of a precision motion system, and then, on the real system. 
\begin{table*}[h]
    \centering
    \caption{GP Hyperparameters for experiments in \cref{sec:case_study}. $c(\task)$ is the constraint limit based on the task, see \eqref{eq:constrained_bo}.}
        \label{tab:gp_hyp}
    \renewcommand{\arraystretch}{1.25}
    \resizebox{0.75\textwidth}{!}{
        \begin{tabular}{@{}c c c@{}}
        \toprule
        Numerical experiment & $f$ & $q$ \\ 
        \cmidrule(l{5mm}r{5mm}){1-3}
        $l = [\Pkp, \Vkp, \Vki, \Aff, \text{stepsize}]^T$ & $[0.4, 0.8, 1.2, 0.3, 0.3]^T$ & $[0.4, 0.8, 1.2, 0.3, 0.3]^T$\\
        $(\sigma_n,\mu_0,\sigmav)$ & (2.5e-3, 1.8, 0.36) & (0.12, $c(\task)$, 1)\\
        \midrule
        Real experiment & $f$ & $q$ \\ 
        \cmidrule(l{5mm}r{5mm}){1-3}  
        $l = [\Pkp, \Vkp, \Vki, \Aff, \text{stepsize}, \text{payload}]^T$ & $[50, 100, 200, 0.5, 0.3, 1]^T$ & $[50, 100, 200, 0.5, 0.3, 1]^T$\\
        $(\sigma_n,\mu_0,\sigmav)$ & (1e-2, 1.8, 0.36) & (9e-2, $c(\task)$, 1)\\
        \bottomrule
        \end{tabular}%
    }

\end{table*}
\subsection{Implementation}
The hyperparameters for the GPs in the experiments are given in \cref{tab:gp_hyp}.
For optimization of the acquisition function, all presented implementations of \goose{} use PSO \cite{pso} with initialization of a set of 50 particles across $\SafeSet$ with a fixed initial velocity of the particles $v_0$, calculated from the $GP^f$ lengthscales (see \eqref{eq:squared_exponential_kernel}) in random directions. In the case of the presented \goose{} implementation of \cite{koenig2021}, which we from now on term as \emph{baseline \goose{}}, the particles expand into the optimistic safe set $\OptimisticSafeSet$ through the PSO update rules, avoiding explicit calculation.
In the case of the presented modified \goose{} implementation, the optimistic safe set $\OptimisticSafeSet$ does not have to be considered, since no expanders are used.

The parameter $\beta = 3$ was fixed for every GP of the implementation, resulting in a $99\%$ confidence interval. For the baseline \goose{} the exploration threshold, determining which parameters of the safe set periphery are defined as expanders, was set to $\textstyle{\epsilon} = 6\sigma_q$ for all experiments.

For the calculation of the final optimum after the experiment, the acquisition function was set to 
\begin{equation}
    \suggestion = \argmin_{x \in \SafeSet} \ucb^f_t(x),
\end{equation}
resulting in the pessimistic optimum from the passive phase.

\begin{figure}[h]
\centering
  \begin{subfigure}[h]{0.95\columnwidth}
    \centering
    \includegraphics[width=\linewidth]{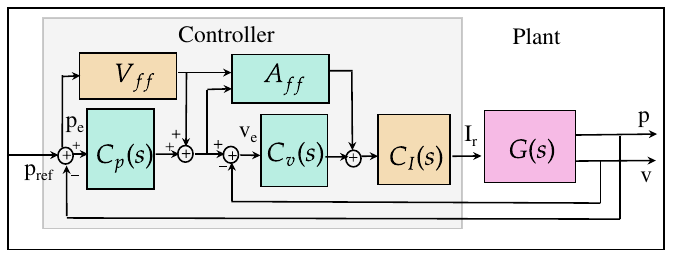}
    \label{fig:Scheme} 
    \end{subfigure}
  \begin{subfigure}[h]{0.7\columnwidth}
  \centering
    \includegraphics[width= 1\linewidth]{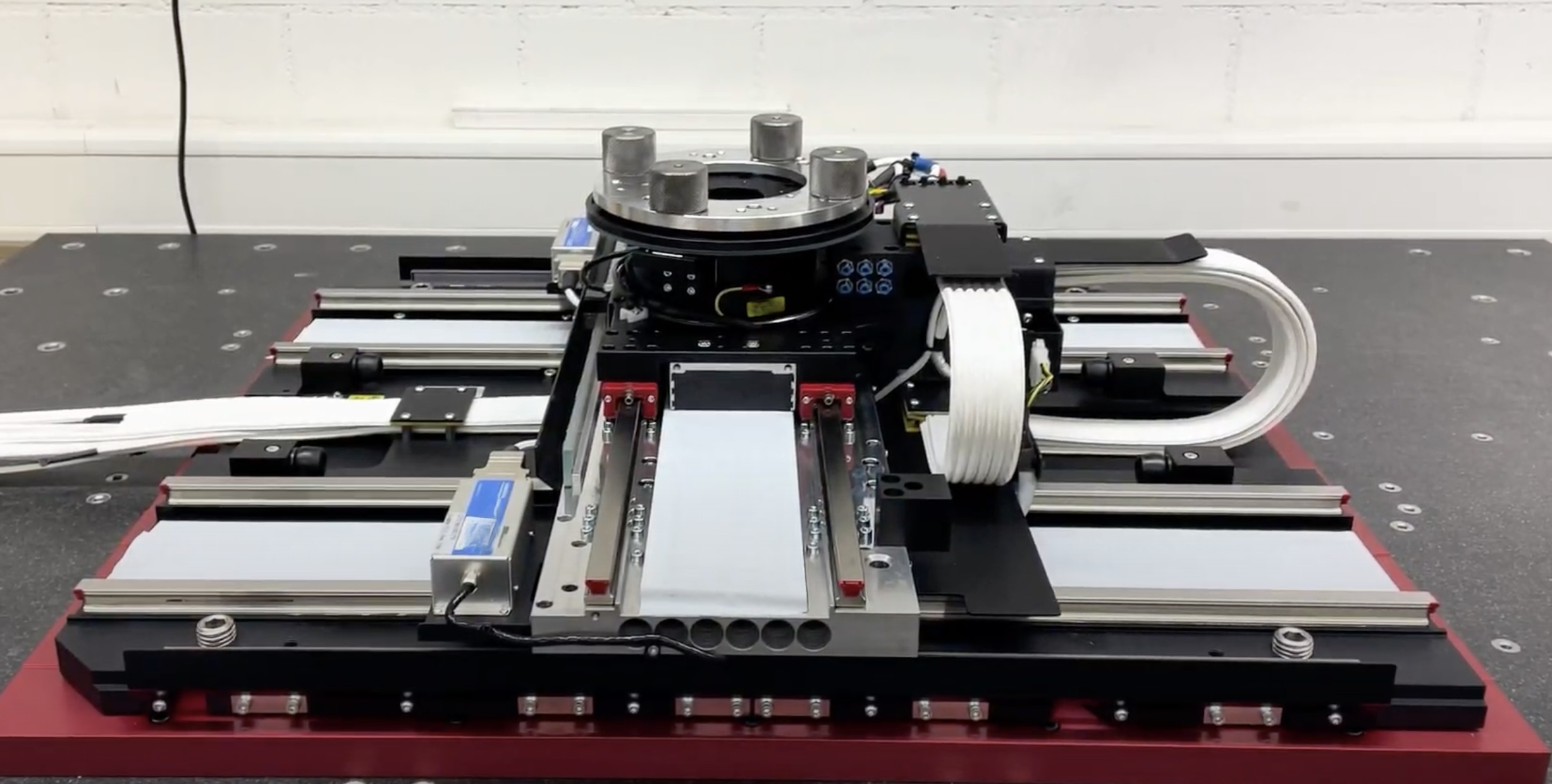}
  \end{subfigure}
  \caption{Top panel: Simplified controller architecture in the experimental study. Top panel. The cyan blocks contain control parameters optimized in this work. Bottom panel: Positioning system by Schneeberger Linear Technology AG.}
  \label{fig:argussetting}
\end{figure}

\subsection{System, controller, and optimization setup}

The system used for experiments of the modified \goose{} algorithm is a linear axis within a high-precision positioning system manufactured by Schneeberger Linear Technology, as illustrated in \cref{fig:argussetting}. This axis is mobilized by a permanent magnet AC motor, outfitted with encoders offering nanometer-level precision for both position and speed monitoring. The axis is guided along two rails using monorail bearings, resulting in positioning accuracy of $\leq 10,\mathrm{\mu m}$, a repeatability margin of $\leq 0.7 ,\mathrm{\mu m}$, and a $3\sigma$ stability of $\leq 1 ,\mathrm{nm}$.
The controller's sampling time, along with the data acquisition rate utilized for this system, stands at $f_s = 20\mathrm{kHz}$.
Such systems find routine application in the semiconductor industry, biomedical engineering, the fields of photonics, and solar technologies, predominantly for production and quality control purposes.

The control system under consideration is a three-level cascade controller, as depicted in \cref{fig:argussetting}. The hierarchical control structure consists of an outermost loop responsible for position control, employing a P-controller represented as $\textstyle{\Cp(s)=\Pkp}$, and a middle loop dedicated to velocity control, employing a PI-controller denoted by $\textstyle{\Cv(s) = \Vkp + \Vki/s}$.
It's worth noting that the inner loop, which governs the current of the drive, is considered well-tuned and remains unchanged throughout the tuning and adaptation process. To expedite the system's response, feedforward structures are employed.
The gain of the velocity feedforward, denoted as $\Vff$, is also considered well-calibrated and remains unaltered during the retuning and adaptation procedure. Conversely, the gain associated with acceleration feedforward, $\Aff$, is subject to adjustment via an algorithm during real system experiments and is subsequently set to $\Aff = 0$ for the simulation experiments.

The cost $f(\mathrm{\bm{x}})$ and constraint $q(\mathrm{\bm{x}})$ are defined as follows:
\begin{equation}
    \begin{split}
        f(\mathrm{\bm{x}}) &= \frac{1}{n_{P} - n_{s}} \textstyle{\sum_{i=n_{s}}^{n_{P}} |\xi(i, n_{s})\mathrm{p_e}(t_{i})|},\\
        q(x) &= \max |\mathcal{F}\left[\xi(i, n_{s})\mathrm{v_e}(t_{i})\right](x, f=f_1)|,\\
        f_1 &=[140\text{Hz}, 1250\text{Hz}],\\
        \xi(i, n_{s}) &= 1 - \frac{1}{1+\mathrm{exp}(-(i-n_{s}-150)/10)},\\
    \end{split}
\end{equation}
where $\mathrm{p_e}(t_{i})$ and $\mathrm{v_e}(t_{i})$ are the position and velocity error from the reference at time instance $t_{i}$ and the settling phase extends from the time index $n_{s}$ to the time index $n_{P}$. The function $\xi(i, n_{s})$ is a left-sided sigmoid function used as a filter for the raw data (in this case, the velocity error). The constraint $q(x)$  is calculated from the Fourier transform of the filtered velocity error. For more details on the performance metrics' definition, we refer the reader to \cite{koenig2023risk}.
The parameters $\mathrm{x}$ that are adapted are the controller gains $\Pkp, \Vkp$, and $\Vki$, as well as the acceleration feedforward gain $\Aff$, and the adaptation is performed  according to the $\log10(\mathrm{stepsize})$ of the motion and (on the experiments on the real system) the payload added to the system, which are also provided as components of the vector $\mathrm{x}$ serving as task parameters (thus, they serve as inputs to the corresponding GPs, but are not optimization variables). The maximum speed, acceleration, and jerk of the reference movement are set to $[v_\text{max}, a_\text{max}, j_\text{max}]^T = [0.9\frac{m}{s}, 20\frac{m}{s^2}, 200\frac{m}{s^3}]^T$.

\subsection{Numerical study}
In a simulation, we demonstrate the adaptability and optimality of the modified \goose{} algorithm. To achieve this, an artificial drift term is added to the output position $p$ in \cref{fig:argussetting}. This drift term is treated as a task parameter in the algorithm, alongside the stepsize. The artificial drift term is increased linearly over the iterations to test the adaptability of the algorithm.
\begin{figure}[h]
\centering
  \begin{subfigure}[h]{1.0\columnwidth}
    \centering    \includegraphics[width=0.7\textwidth]{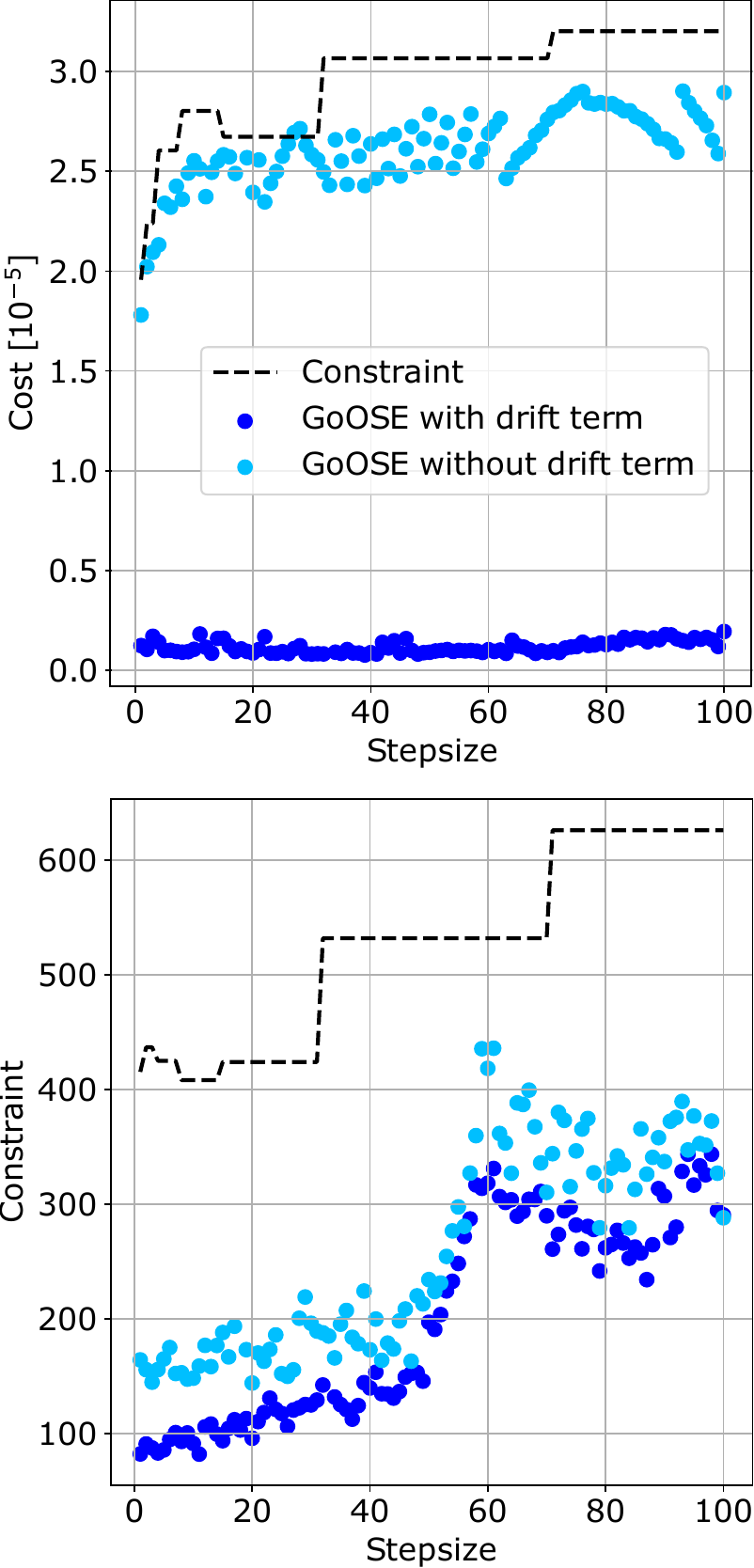}
  \end{subfigure}
    \caption{Simulation: Comparison of the modified \goose{} with one accounting for the drift term as a task parameter and the other without. The plots show the achieved cost and constraint value with respect to stepsize after convergence of the algorithm. It is evident that the \goose{} without the drift term has difficulty achieving optimality due to the presence of the artificial drift.}    
    \label{fig:GoOSE_w_vs_wo_drift} 
\end{figure}

\begin{figure}[h]
\centering
  \begin{subfigure}[h]{1.0\columnwidth}
    \centering    \includegraphics[width=\textwidth]{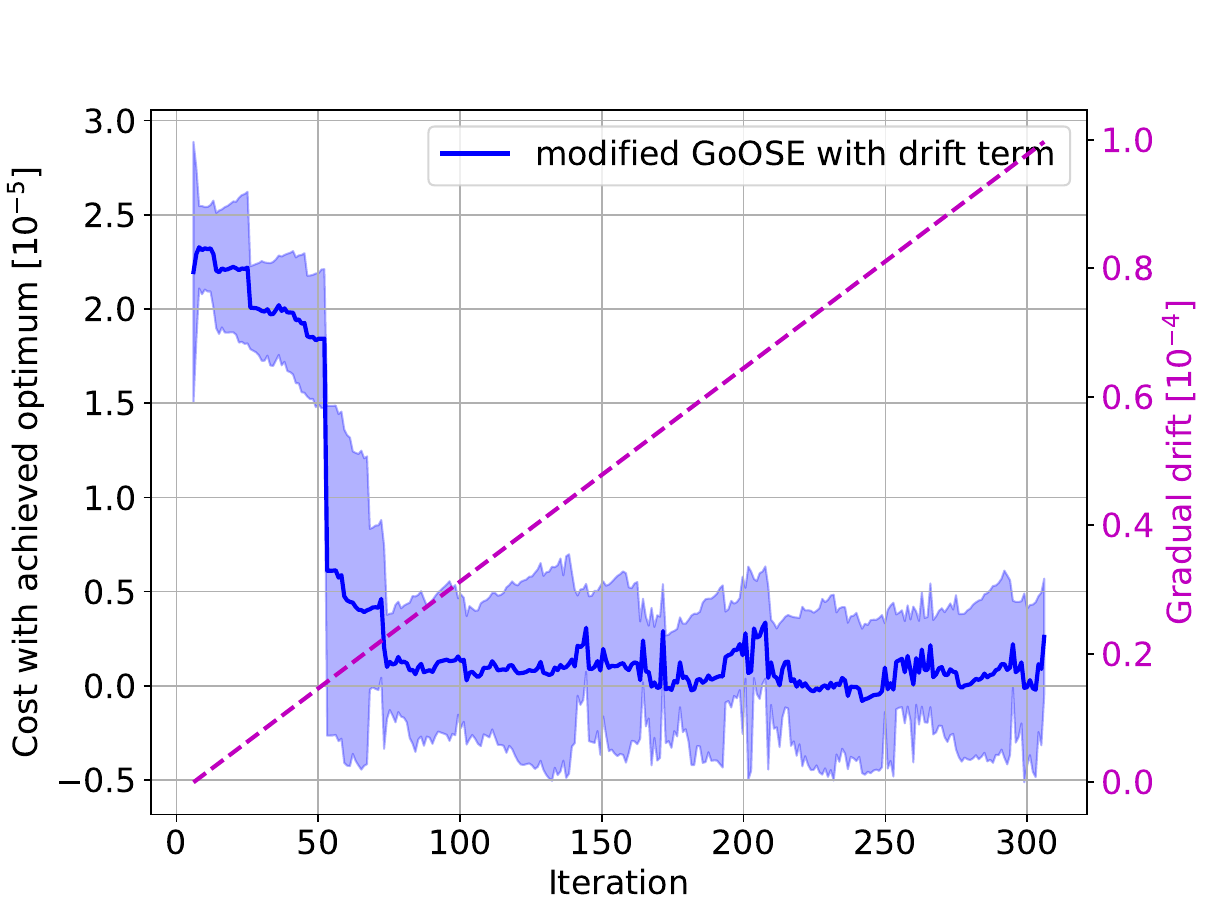}
  \end{subfigure}
    \caption{Simulation: The achieved optimum cost over the iterations is depicted for the modified \goose{} with the drift term. It is able to compensate for the drift by constantly modifying the controller gains and thus to keep the cost at the optimum.}    
    \label{fig:Optima_over_iterations_simulation} 
\end{figure}

 In \cref{fig:GoOSE_w_vs_wo_drift}, we compare two modified \goose{} algorithms, one with a task dimension for the drift term and one without. The cost values show that the modified \goose{} with the drift term can account for the artificial drift and keep the cost at the optimum. Furthermore, \cref{fig:Optima_over_iterations_simulation} demonstrates how the modified \goose{} algorithm keeps the cost at the minimum despite the drift term by constantly modifying the gains.
 The results show successful adaptation of the proposed method to shifts due to external disturbances. Such adaptation scenarios have practical relevance in cases such as changing die sizes or weights in a semiconductor manufacturing scenario, or changing cutting tools with different weights in a laser cutting systems.

\subsection{Experimental study}
\subsubsection{Comparisons to the baseline \goose{}}
In a first experiment on the real system, we compare the baseline \goose{} with the modified \goose{} algorithm described in \cref{sec:method}. For this, we run both optimizations for $T=135$ iterations with a fixed stepsize of 10mm and a payload that is changed every 15 iterations between 0.4kg and 2kg. Both algorithms start with an identical initial set of data $x_0 = [\Pkp, \Vkp, \Vki, \Aff, \text{log}_{10}(\text{stepsize}), \text{payload}]^T = [200, 600, 1000, [0, 1, 2], 1, 0.4]^T$. The predefined safe set is set to $\SafeSeed = [\Pkp, \Vkp, \Vki, \Aff]^T = [200, 600, 1000, 0]^T$ for both algorithms. 

\begin{figure}
\centering
  \begin{subfigure}[h]{\columnwidth}
    \centering   
    \includegraphics[width=\textwidth,trim={0.75cm 0.1cm 0cm 0cm},clip]{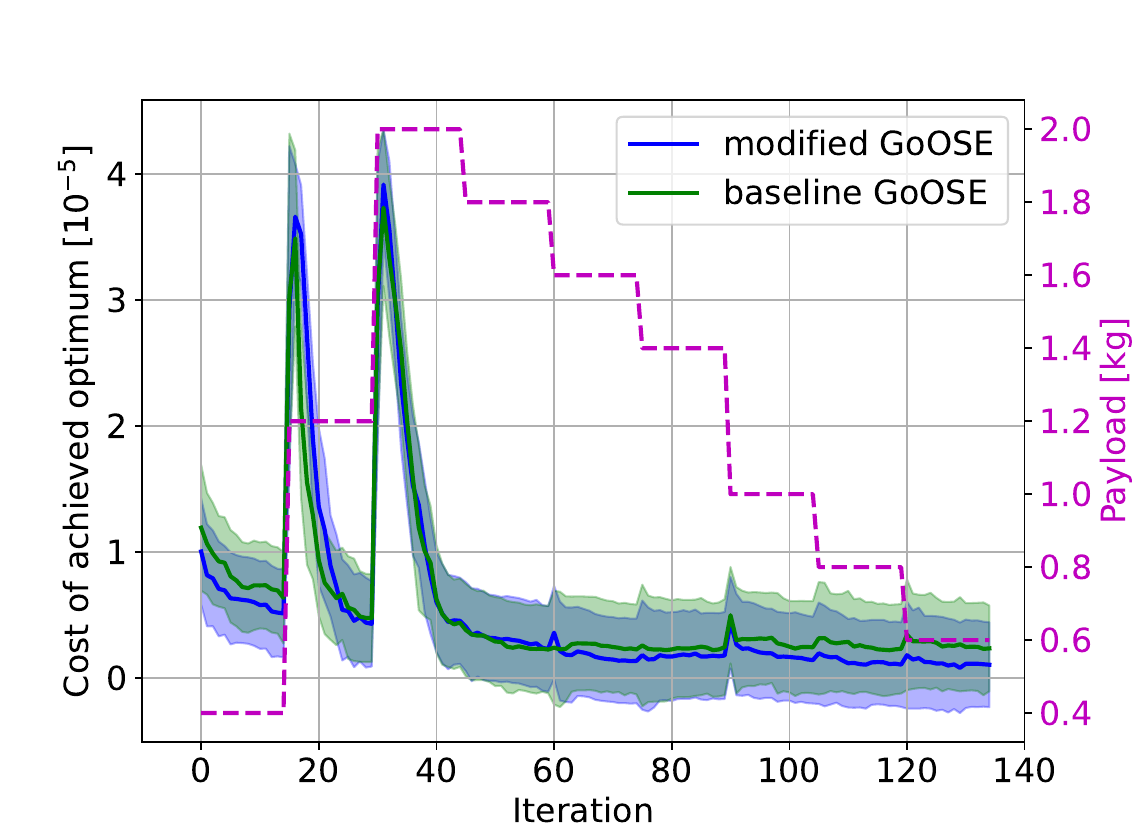}
    \caption{Comparison of the rolling optimum. The solid line shows the prediction of the optimum at every iteration the lighter areas indicate the confidence intervals.}    
    \label{fig:Optima_over_iterations} 
  \end{subfigure}
  \begin{subfigure}[h]{\columnwidth}
   \includegraphics[width=\textwidth,trim={0.75cm 0.1cm 0cm 0cm},clip]{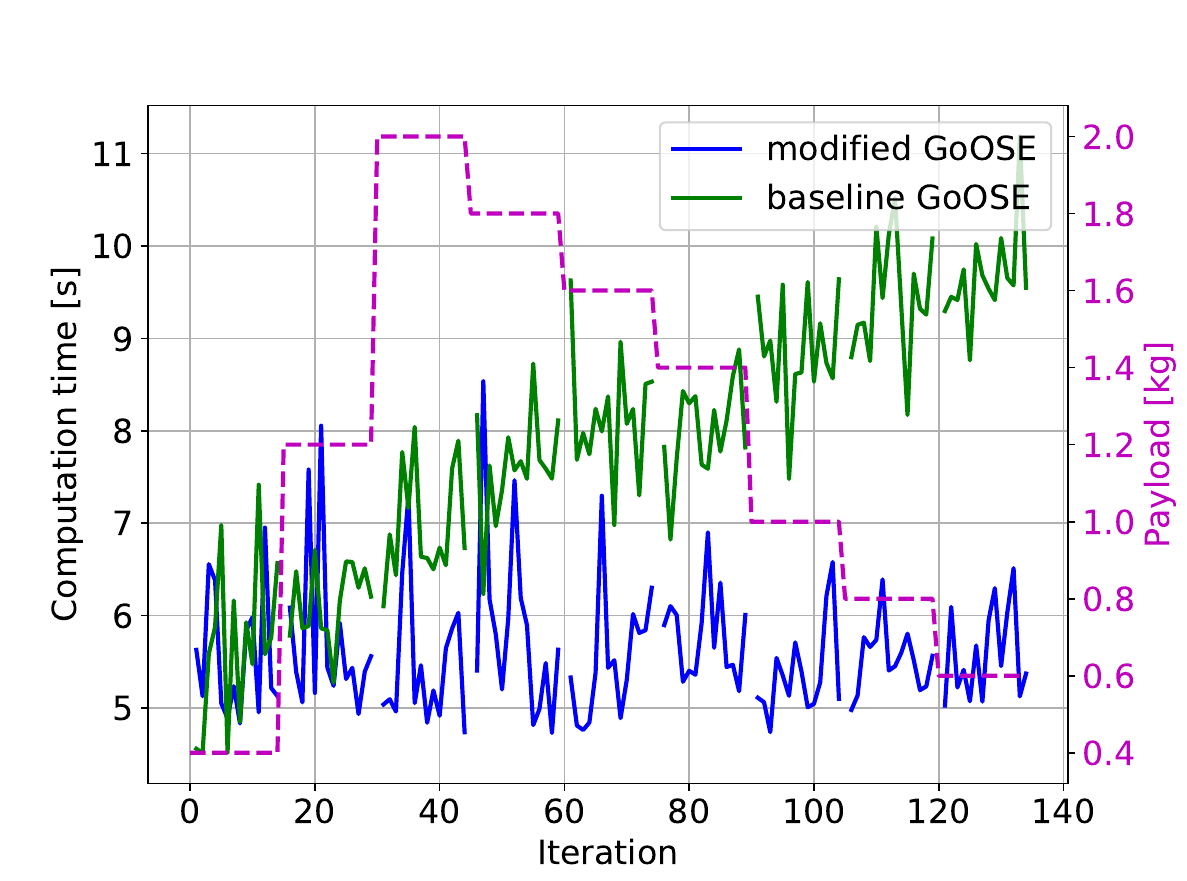}
    \caption{Comparison of the total iteration time including the machine movement of around 2.4s.}  
    \label{fig:Computation_comparison} 
  \end{subfigure}
  \caption{Comparison of the baseline \goose{} with the modified \goose{} tuning executed on the Argus system. The stepsize is fixed to a 10mm movement, while the payload is altered every 15 iterations.}
\end{figure}

\autoref{fig:Optima_over_iterations} shows the convergence of both algorithms after alternation of the payload is comparably fast, while the modified \goose{} performs slightly better, by comparing the rolling optimum calculated during the passive phase.
Furthermore, it is shown that for both algorithms each further alternation of the payload requires fewer iterations to find the global optimum for the given task. After 3 alternations of the payload, the search for the optimum finishes almost instantaneously. Comparing the iteration time of both algorithms we can see that the modified \goose{} requires approximately the same computational time for each of the $T=135$ iterations, while the original \goose{} computational time grows with each new data point, see \cref{fig:Computation_comparison}.
The figure shows the complete iteration time of the experiment including the machine's movement of approximately 2.4s per iteration (10mm back and forth movement).

\begin{figure}
\centering
  \begin{subfigure}[h]{1.0\columnwidth}
    \centering  \includegraphics[width=0.65\textwidth]{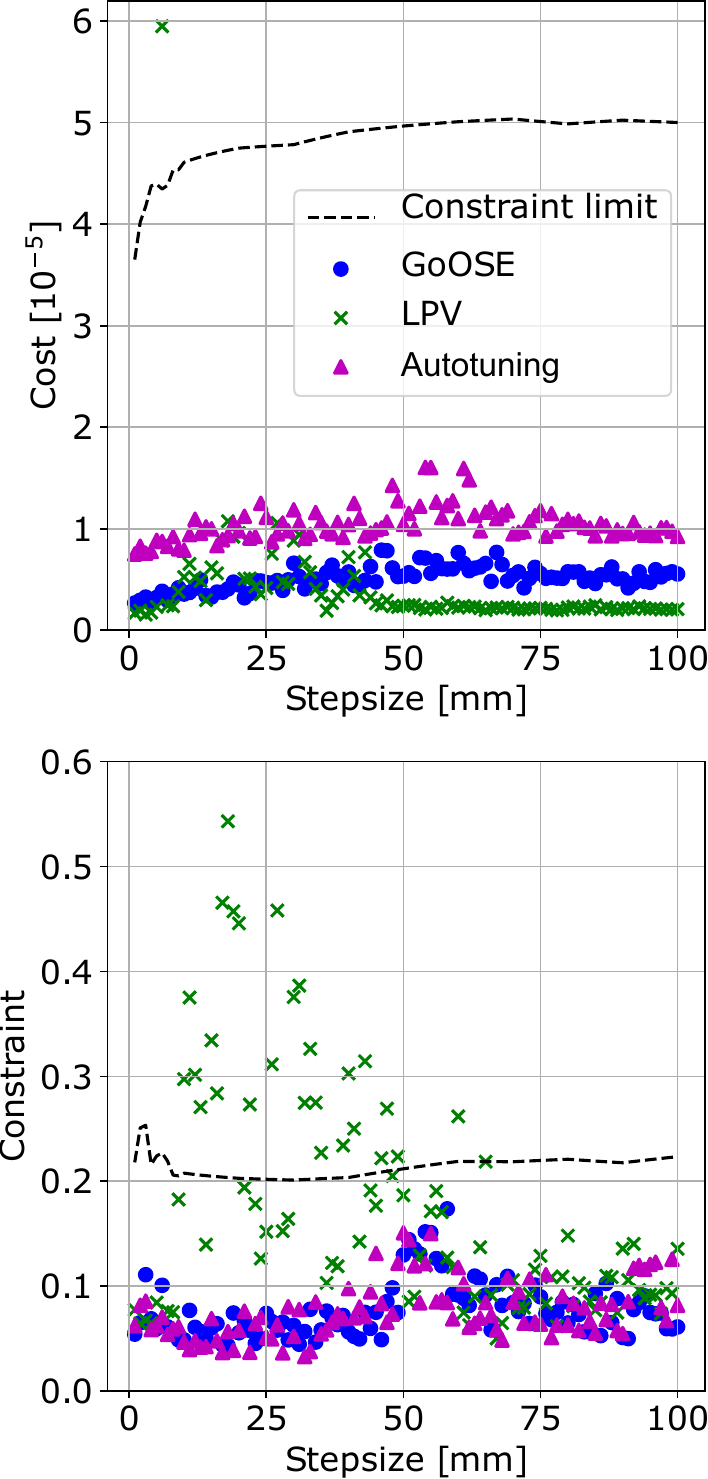}
  \end{subfigure}
  \caption{Comparison of the cost and constraint of the identified optima calculated by the modified \goose{} algorithm, interpolated grid optimization LPV, and the built-in autotuning heuristic of the machine. In the case of \goose{} and interpolated grid optimization LPV, the optimum depends on the two adaptive task parameters defined as the stepsize and the payload. For illustration, the payload in the figure is fixed to 0.4kg and the x-axis displays the stepsize.}
\label{fig:GoOSE_vs_LPV_ACS}
\end{figure}

We conduct a second experiment on the real system to obtain the cost and constraint optima of the modified \goose{} (see \cref{fig:GoOSE_vs_LPV_ACS}). 
We use the initial samples 
\begin{align*}
    x_0 &= [\Pkp, \Vkp, \Vki, \Aff, \text{log}_{10}(\text{stepsize}), \text{payload}]^T, \\
    &= [200, 600, 1000, [0, 1, 2], [0, 2], 0.4]^T,
\end{align*}
predefined safe set $\SafeSeed = [\Pkp, \Vkp, \Vki, \Aff]^T = [200, 600, 1000, 0]^T$, $T=300$ iterations with random step and alteration of the payload every 33 iterations between 0.4kg and 2kg, is compared with the optima of the built heuristic non adaptive autotuning of the machine and an adaptive interpolated grid optimization LPV, where a grid of 1900 candidates $x \in X$ is recorded (all permutations of $\optimizable = [[200, 300, 400], [600, 800, 1000], [1000, 1250, 1500]]^T$ are combined with all permutations of $\task = [\text{log}_{10}([1,2,5,10,20,50,100]),[0.4, 1.2]]^T$. 

The optimum for all recorded tasks fulfilling the constraint is calculated. For a task in between or beyond known tasks, a linear interpolation strategy is chosen.
It is shown that the modified \goose{} shows better performance than the heuristics-based autotuning of the controller. Both modified \goose{} and autotuning never violate the constraint and keep the system stable for any task, while the interpolated grid optimization LPV often violates the stability constraint and also drives the system once into an unstable position error for the 6mm step, where the automatic emergency stop was triggered, even though the interpolated grid optimization LPV was specifically trained for the shown payload of 0.4kg, and the other two approaches are adaptive also with respect to payload (thus, not specifically tuned to this payload). This can be explained by two effects. 1) The 6mm step was not directly tuned during the grid optimization but interpolated. 2) The interpolated grid optimization LPV method is not robust enough w.r.t. disturbances in the system, while \goose{} handles the disturbances due to the noise model in the GP model (see $\sigma_n$ in \eqref{eq:GP_posterior} and \cref{tab:gp_hyp}).

\subsubsection{Comparison of parallel schemes}
\begin{figure}
\centering
  \begin{subfigure}[h]{1.0\columnwidth}
    \centering    \includegraphics[width=\textwidth,trim={0.25cm 0cm 0cm 0cm},clip]{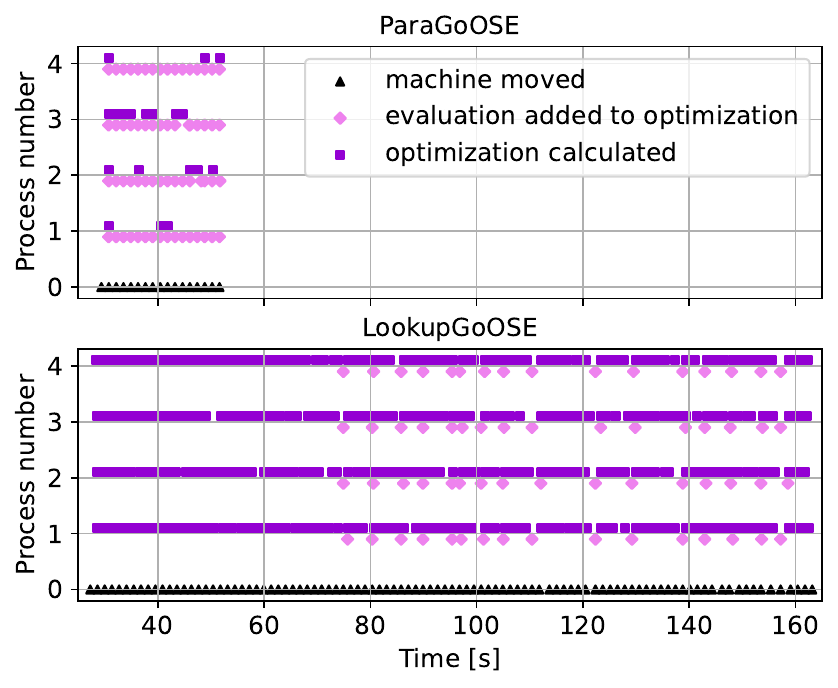}
  \end{subfigure}
    \caption{ParaGoOSE vs LookupGoOSE: First active phase of both algorithms, both algorithms are stopped after the first active phase is completed. The termination criterion is set to adding a total of 16 new data points to the GPs.
    Both algorithms work independently of the machine scheduling and do not prohibit the machine from moving, due to parallelization.
    It can be seen that Para\goose{} is significantly faster at acquiring the data points, due to much fewer measurements being ignored.
    Furthermore, in the first ~$75$ seconds of the Lookup\goose{}, no data is added to the GPs, because the optimization processes have to update the whole task grid.}    
    \label{fig:Parralel_goose_comparison} 
\end{figure}

In a final experiment on the real system the two parallel schemes for the adaptive control modified \goose{} algorithm, described in \cref{sec:method} are compared. We create a similar experiment as used for \cref{fig:GoOSE_vs_LPV_ACS}. Here the tuning contains random stepsizes and 3 different payload settings altering between 0.4kg, 1.2kg, and 2.0kg. The initial samples $x_0$ and predefined safe set $\SafeSeed$ are equivalent to the previous experiment. Both parallel adaptive control algorithms use 4 parallel processes for optimization. 
While using Para\goose{} the predefined safe set $\SafeSeed$ is used on the machine, if the optimization is not fast enough to deliver the corresponding optimizer/optimum (e.g. after the start of the algorithm).
Similarly while using Lookup\goose{} the grid of optimizers/optima is initialized using the predefined safe set $\SafeSeed$ as the optimizer for every entry in the task grid. The task grid is selected as all permutations of the task 
\begin{align*}
    \X_{\task} &= [\text{log}_{10}(\text{stepsize}), \text{payload}]^T, \\
    &= [\text{log}_{10}([1,2, ...,10,20,...,100]), [0.4,0.6,...,2.0]]^T.
\end{align*}
As discretization parameter of the neighborhood $\Delta\task = [0.3, 0.2]^T$ is chosen. \cref{fig:Parralel_goose_comparison} summarizes the performance of both parallel schemes. The advantage of the Para\goose{} is evident by looking at the number of added evaluations to the GPs since all the future tasks were known in advance for this experiment. Otherwise, the Lookup\goose{} can handle the unknown future tasks with the bottleneck of its speed due to having to calculate the optima for a neighborhood of the applied task.

\section{Conclusion}
\label{sec:conclusion}
In this work, we present two run-time online Bayesian optimization algorithms that safely perform constrained optimization. 
To achieve continuous adaptation, we parallelize optimization iterations and further reduce the computation time of the \goose{} algorithm, by removing the expensive expander calculation of prior \goose{} implementations, while maintaining performance and safety. 
We demonstrate the proposed framework on a controller tuning and run-time add-on adaptive control application where high performance and safe evaluations are crucial to ensure physical safety without disruption of the operation.
We present results in both simulations and on a real high-precision motion system to show the improved performance over state-of-the-art tuning and adaptive control methods that can handle arbitrary controller structures. 

Our work was motivated by the adaptation of fast controllers in time-critical applications, however, the presented framework can be effective on a variety of control tasks with various controller structures. Future work will study the implementation of the proposed method in safety-critical industrial robotics applications.

\bibliography{bib,report_bibs}
\bibliographystyle{IEEEtran}


\end{document}